\documentclass[
  twocolumn,
  prb,
  showpacs,
  amsmath,
  amssymb,
  superscriptaddress,
  floatfix,
  groupedaddress
]{revtex4-1}
\usepackage[utf8]{inputenc}
\usepackage{amsmath,amssymb,graphicx}
\bibpunct{[}{]}{,}{n}{}{}
\begin{document}
\title{Effect of flexural phonons on the hole states in single-layer black phosphorus}
\author{S. Brener}
\author{A.~N. Rudenko}
\email[]{a.rudenko@science.ru.nl}
\author{M.~I. Katsnelson}
\affiliation{\mbox{Radboud University, Institute for Molecules and Materials, Heijendaalseweg 135, 6525 AJ Nijmegen, The Netherlands}}
\date{\today}
\begin{abstract}
Flexural thermal fluctuations in crystalline membranes affect the band structure of the carriers, which leads to an exponential density-of-states (DOS) tail beyond the unperturbed band edge. We present a theoretical description of this tail for a particular case of holes in single-layer black phosphorus, a material which exhibits an extremely anisotropic quasi-one-dimensional dispersion ($m_y/m_x\gg1$) and, as a result, an enhanced Van Hove singularity at the valence band top. The material parameters are determined by {\it ab initio} calculations and then are used for quantitative estimation of the effect of two-phonon (flexural) processes have on the charge carrier DOS. It is shown that unlike the isotropic case, the physics is determined by the phonons with wavevectors of the order of $q^*$, where
$q^*$ determines the crossover between harmonic and anharmonic behavior of the flexural phonons. The spectral density of the holes in single-layer black phosphorus at finite temperatures is calculated.
\end{abstract}

\maketitle

Few-layer black phosphorus attracts a lot of attention now as a prospective two-dimensional (2D) semiconducting material with a tunable energy gap \cite{Li2014,Liu2014,Qiao2014,Xia2014,Ling2015}. Being, as other 2D materials, an atomic-thick crystalline membrane, it is affected by thermal flexural fluctuations which results in anomalous structural and mechanical properties at finite temperatures \cite{nelson,AL,AN,Doussal,book-Nelson,book-Katsnelson,ACR,nelson16,mech}. It was suggested \cite{flexuron} that these thermal fluctuations can essentially influence on the states of charge carriers in 2D semiconductors and even lead to an autolocalization, that is, formation of a self-trapped state (flexuron). That general consideration assumed an isotropic electronic spectrum whereas black phosphorus is highly anisotropic, with dramatically different effective masses, $m_y/m_x\gg1$. Especially, in the hole-doped case the situation is quite peculiar, with a formation of Van Hove singularity near the top of the valence band which is essentially enhanced by mass anisotropy \cite{vanhove}. One can expect that the holes in single-layer black phosphorus (also called phosphorene) are quasi-one-dimensional quasiparticles and thus are strong candidates for a pronounced effect of localization due to interaction with bosonic modes.

In general, deformation of the lattice in 2D materials results in distortion of the carrier dispersion. For different types of deformations this effect can be accounted for by different phonon modes interacting with fermionic charge carriers. The interaction parameters then can be determined either from experiment or from different theoretical approaches. In 2D materials like graphene or black phosphorus it is a well-justified approach to separate in-plane from out-of-plane, or flexural phonon modes for not too small wavevectors \cite{ACR}. On the other hand, for smaller wavevectors these modes begin to interact which results in changing of the functional form of the correlation function dependence on the wavevector \cite{Doussal,book-Nelson,book-Katsnelson,ACR}. This will be discussed in more detail below, in particular for the anisotropic case.

In what follows we calculate the spectral function and the density of hole states (DOS) in black phosphorus renormalized due to the interaction with flexural phonons. To this end, we use the 
diagrammatic approach and calculate the hole self-energy using the Ward identity as a low momentum transfer approximation for the vertex. This seems to be the minimal model that allows one to get 
an exponential DOS tail and is thus a more suitable approach than e.g. the self-consistent Born approximation. This approximation was earlier suggested in a context of magnetic semiconductors near the critical point \cite{tail}. Its relation to the path-integral approach used in Ref.~\onlinecite{flexuron} is discussed in Ref.~\onlinecite{fluctuon}. 

The fermionic Hamiltonian is given by
\begin{equation}
    \hat{H}_e=\sum_{\mathbf p}\epsilon_{\mathbf p}c^\dagger_{\mathbf p}c_{\mathbf p}+\int  d{\mathbf r} V^{\mathrm{eff}}(\mathbf r)c^\dagger_{\mathbf r}c_{\mathbf r}.
\end{equation}
Here $c^\dagger_{\mathbf {p,r}}$, $c_{\mathbf {p,r}}$ are the creation and annihilation operators of the fermions with momentum $\mathbf p$ (at point $\mathbf r$), $\epsilon_{\mathbf p}$ is the dispersion of the fermions and $V^{\mathrm{eff}}(\mathbf r)$ is the effective potential acting on them due to lattice distortion. For flexural deformations in systems with symmetry plane (or gliding plane as in case of black phosphorus), the effective potential is quadratic in gradients of the lattice atoms shift $h(\mathbf r)$ in the direction, perpendicular to the sample plane $(x,y)$ \cite{flexuron,one-phonon}:
\begin{eqnarray}
\label{potential}
\notag
    V^{\mathrm{eff}}(\mathbf r)=g_x\left(\frac{\partial h}{\partial x}\right)^2+g_y\left(\frac{\partial h}{\partial y}\right)^2 = \quad \quad \quad \quad \\ 
= \sum_{\mathbf{q_1,q_2}}h_{\mathbf{q_1}}h_{\mathbf{q_2}}(g_xq_{1x}q_{2x}+g_yq_{1y}q_{2y})e^{i\mathbf{r(q_1+q_2)}}.
\end{eqnarray}
The mixed $xy$ term is suppressed for the orthorhombic symmetry and we disregard it in our consideration. $h_{\mathbf q}$ is the Fourier transform of the flexural lattice displacement field.

Correlators of the $h$ field are found from the elasticity theory equations for membranes. For isotropic membranes the well-known result is \cite{Doussal,book-Nelson,book-Katsnelson,ACR,flexuron}
\begin{equation}
\label{cumulant_isotropic}
    \left<h_{q}h_{-q}\right>=\left\{
    \begin{array}{lcc}
         \frac{T}{\varkappa q^4}&\mbox{for}&q>q^*  \\
          \frac{T}{\varkappa q^{4-\eta}q^{*\eta}}&\mbox{for}&q<q^*
    \end{array}
    \right.
\end{equation}
Here $\left<\dots\right>$ denotes Gibbs averaging, $T$ is the temperature, $\varkappa$ is the material constant, determining the flexural rigidity of the membrane, $\eta$ is the critical exponent ($\eta\approx0.82$ in the self-consistent screening approximation \cite{Doussal} and $\eta\approx0.85$ in Monte Carlo simulations for graphene \cite{ACR}) and $q^*$ is the characteristic wavevector, such that for $q<q^*$ the interplay between flexural and in-plane deformations becomes crucial and softens the flexural modes. Neglecting this effect would lead to unphysical low wavevector divergence in $\left<(V^{\mathrm{eff}})^2\right>$. In the isotropic case $q^*$ is given by
\begin{equation}
    q^*=\sqrt{\frac{3TY}{16\pi\varkappa^2}},
\end{equation}
where $Y$ is the Young modulus \cite{nelson}.

Before we proceed, note that averaging of $V^{\mathrm{eff}}$ yields a constant that just shifts the whole fermionic band and has no non-trivial effects. To account for those we have to calculate the second cumulant of the bosonic field ${\mathcal K}_2(\mathbf q)=\left<(V^{\mathrm{eff}})^2\right>_{\mathbf q}-\left<(V^{\mathrm{eff}})\right>_{\bf q}^2$. Then, neglecting higher cumulants, we can write for the fermionic self-energy
\begin{equation}
    \Sigma(E,\mathbf p)=\int\frac{d^2q}{(2\pi)^2}\gamma(\mathbf{p-q,p,q},E){\mathcal K}_2(\mathbf q)G(E,\mathbf{p-q})),
\end{equation}
where $G$ is the fermion's Green's function, $\gamma$ is the three-leg vertex and $E$ is the energy of the fermion (cf. Refs.~\onlinecite{tail,fluctuon}). The approximation $\gamma=1$ corresponds to the self-consistent Born approximation which is supposed to be enough for positive energies (that is, within the band). To take into account also the fluctuation DOS tail, vertex corrections should be taken into account. At this stage we argue that for energies far enough from the band edge only small momentum transfers play a role \cite{tail,fluctuon}, so that we neglect momentum dependence of the vertex and of the self-energy and use the Ward identity for the vertex:
\begin{equation}
    \gamma(\mathbf{p,p},0,E)=1-\frac{\partial\Sigma}{\partial E},
\end{equation}
ending up with the following closed differential equation for the self-energy:
\begin{equation}
\label{sigma_equation}
    \Sigma(E)=\left(1-\frac{\partial\Sigma}{\partial E}\right)\int \frac{d^2q}{(2\pi)^2}\frac{{\mathcal K}_2(\mathbf q)}{E-\epsilon_{\mathbf q}-\Sigma(E)}.
\end{equation}

Now we go on to estimate ${\mathcal K}_2(\mathbf q)$ for the anisotropic case, which is a question of its own significance. For $q>q^*$ it has been done recently in Ref.~\onlinecite{mobility}, so here we do the calculations for small wavevectors.

A consistent theory of anisotropic membranes is to our knowledge not yet fully developed, so in this paper we make plausible assumptions for the low-$q$ behavior of the cumulant $\left<h_{\mathbf q}h_{\mathbf {-q}}\right>$ providing their consistence with Eq.~(\ref{cumulant_isotropic}).

Following the reasoning of Ref.~\onlinecite{mobility}, we model the anisotropy of the bending rigidity by writing for large $q$:
\begin{equation}
   \left<h_{\mathbf q}h_{\mathbf {-q}}\right>= \frac{T}{\left(\varkappa_x^{1/2} q_x^2+\varkappa_y^{1/2} q_y^2\right)^2}.
\end{equation}
As a next step, we rescale the $x,y$-components of the wavevector introducing $q'_{x,y}=q_{x,y}\varkappa_{x,y}^{1/4}$, and write
\begin{equation}
\label{cumulant_anisotropic}
    \left<h_{\mathbf q}h_{\mathbf{-q}}\right>=\left\{
    \begin{array}{lcc}
         \frac{T}{q'^4}&\mbox{for}&q'>q'^*  \\
          \frac{T}{q'^{4-\eta}q'^{*\eta}}&\mbox{for}&q'<q'^*
    \end{array}
    \right.
\end{equation}
with $q'=(q'^2_x+q'^2_y)^{1/2}$, $(q'^*)^2=3T(Y_xY_y)^{1/2}/(16\pi(\varkappa_x\varkappa_y)^{3/4})$ and $Y_{x,y}$ is the Young moduli in $x,y$ directions. Finally, introducing $a_{x,y}=g_{x,y}\varkappa_{x,y}^{-1/2}$ and shifting $\mathbf k'\rightarrow\mathbf{k'-q'/2}$, we obtain for ${\mathcal K}_2(\mathbf q)$:
\begin{eqnarray}
\notag
{\mathcal K}_2(\mathbf q)= \frac{2}{(\varkappa_x\varkappa_y)^{1/4}}\int\frac{d^2k'}{(2\pi)^2} \left [a_x\left(k_x'^2-\frac{q_x'^2}{4}\right) + \right. \quad \quad \\
\left. + a_y\left(k_y'^2-\frac{q_y'^2}{4}\right)\right]^2\left<h_{\mathbf{k}+\frac{\mathbf{q}}{2}}h_{-\mathbf{k}-\frac{\mathbf{q}}{2}}\right> \left<h_{\frac{\mathbf{q}}{2}-\mathbf{k}}h_{\mathbf{k}-\frac{\mathbf{q}}{2}}\right>. \quad
\end{eqnarray}
To get this expression, we take $V^{\mathrm{eff}}$ given by Eq.~(\ref{potential}) and calculate the Fourier transform of $\left<V^{\mathrm{eff}}(\mathbf{r}_1)V^{\mathrm{eff}}(\mathbf{r}_2)\right>$ over $\mathbf{r}_1-\mathbf{r}_2$, which is done using the Wick's theorem to decouple the average of the product of four $h_{\mathbf q}$'s. The latter corresponds to the self-consistent screening approximation \cite{Doussal}. Finally, the rescaling of the components of all wavevectors is made as described above.

As the conditions $q'<q'^*$, $q>q'^*$ are rather crossovers than strict boundaries, we can within the same accuracy split the $\mathbf k'$-integral into two parts (1) $k'<q'^*$ and (2) $k'>q'^*$. For the first one we take the anharmonic expression for the $\left<h_{\bf q}h_{-{\bf q}}\right>$ correlators, and for the second part we use the harmonic one. Moreover, we limit ourselves to calculating only the leading terms in $q'/q'^*$, so for $k'>q'^*$ we just neglect the $q'$-dependence to obtain the contribution to $\mathcal{K}_2$:
\begin{eqnarray}
\label{bigk}
\notag
{\mathcal K}_2(\mathbf q)\longleftarrow  \frac{2T^2}{(\varkappa_x\varkappa_y)^{1/4}} \int_{k'>q'^*}\frac{d^2k'}{(2\pi)^2}\frac{\left(a_xk_{x}'^{2}+a_yk_{y}'^{2}\right)^2}{k'^8}= \\
=\frac{T^2(3a_x^2+2a_xa_y+3a_y^2)}{16\pi(\varkappa_x\varkappa_y)^{1/4}(q'^*)^2} \quad \quad \quad 
\end{eqnarray}
The integration over $k'<q'^*$ is more involved, as we have to keep the $\mathbf q'$-dependence of the integrand, which leads to anisotropic contributions. Here, apart from the leading term $\propto(q'/q'^*)^{2\eta-2}$, we also keep the subleading term $\propto(q'/q'^*)^{0}$, as for $\eta=0.85$ they are numerically of the same order of magnitude for not extremely small $q'$. Evaluating the integral \cite{app} and adding it up with the contribution given by Eq.~(\ref{bigk}), we arrive at
\begin{widetext}
\begin{multline}
\label{smallk}
{\mathcal K}_2(\mathbf q)= \frac{T^24^{1-\eta}}{2\pi(\varkappa_x\varkappa_y)^{1/4}(q'^*)^{2\eta}q'^{2-2\eta}}\times\\
\Bigl(\cos{(4\alpha)}(a_x-a_y)^2\cdot0.062+\cos{(2\alpha)}(a_x^2-a_y^2)\cdot0.54+(a_x+a_y)^2\cdot1.59+(a_x-a_y)^2\cdot1.03\Bigr)-\\
\frac{T^2(3a_x^2+2a_xa_y+3a_y^2)}{16\pi(\varkappa_x\varkappa_y)^{1/4}(q'^*)^2}\frac{\eta}{1-\eta}.
\end{multline}
\end{widetext}
Here the numerical factors in the parentheses are results of numerical integration, for which the specific value for $\eta=0.85$ was used; $\alpha$ is the polar angle of the $\mathbf q'$ vector (not the $\mathbf q$ vector!). 

The case $q'>q'^*$ has been considered in Ref.~\onlinecite{mobility}. The result is:
\begin{multline}
\label{bigq}
{\mathcal K}_2(\mathbf q)=\frac{T^2}{4\pi(\varkappa_x\varkappa_y)^{1/4}q'^{2}}\times \\
\left(\cos{(4\alpha)}\frac{(a_x-a_y)^2}{2}+\cos{(2\alpha)}(a_x^2-a_y^2)\ln{\frac{\gamma q'}{q'^*}}+\right.\\
\left.+(a_x^2+a_y^2)\ln{\frac{\gamma q'}{q'^*}}+\frac{(a_x+a_y)^2}{4}\right),
\end{multline}
where $\gamma\gtrsim1$ is some number determined by details of $\langle h_qh_{-q}\rangle$ behavior for $q\sim q^*$.

After obtaining the above expressions for $\mathcal{K}_2$, our immediate goal is to evaluate the integral in Eq.~(\ref{sigma_equation}). As it has been mentioned above, the dispersion of the holes in black phosphorus is quasi-one-dimensional. It has been evaluated using {\it ab initio} numerical methods in our previous work \cite{mobility} and can be approximated in atomic units by
\begin{equation}
\epsilon_{\mathbf q}=2.9q_x^2+0.05q_y^2+13.2q_y^4.
\end{equation}
In these units, the averaged $q^*\approx0.015$ for room temperature. Note that if the coefficient at $q_y^2$ were exactly zero we would have an enhanced Van Hove singularity in the bare density of states $\propto E^{-1/4}$ at the band edge, instead of a step in generic 2D case. 
 
As it will be shown below, the integral in Eq.~(\ref{sigma_equation}) is determined by $q\sim q^*$. For these values the first term in the dispersion dominates, so we can take
\begin{equation}
\epsilon_{\mathbf q}=\frac{q_x^2}{2m_x}.
\end{equation}
We perform the same variable change $\mathbf q\rightarrow\mathbf q'$ and introduce $\varepsilon=-E+\Sigma(E)>0$ to rewrite our equation as
\begin{equation}
\label{sigma_2}
\varepsilon+E=\frac{\partial\varepsilon}{\partial E}\int_0^{\infty}\int_0^{2\pi}\frac{q'dq'd\alpha}{(2\pi)^2(\varkappa_x\varkappa_y)^{1/4}} \frac{\mathcal{K}_2(\mathbf q)}{\varepsilon+\frac{q'^2\cos^2{\alpha}}{2m_x\varkappa_x^{1/2}}}.
\end{equation}

Consider first $q'<q'^*$. As
\begin{equation}
\label{cosint}
\int_0^{2\pi}\frac{d\alpha}{1+a\cos^2{\alpha}}=\frac{2\pi}{ \sqrt{1+a}}
\end{equation}
and $\mathcal K_2\propto (q')^{2\eta-2}$ with $\eta=0.85>0.5$, we easily confirm, that for small $\varepsilon$ the $q'$-integral is determined by the upper limit $q'^*$. Analogously, for $q'>q'^*$, the $q'$-integral is determined by the lower limit, as $\mathcal K_2\propto (q')^{-2}$. Note that for conventional 2D-dispersion, this reasoning does not work and the integral acquires its value on $q' \rightarrow 0$ for $\varepsilon<(q^*)^2/2m$.

Estimation of the integral for $E\ll q^2/2m_x$ allows us to recast Eq.~(\ref{sigma_2}) as
\begin{equation}
\label{eqsigma}
\varepsilon+E=\frac{\partial\varepsilon}{\partial E}\frac{(T\beta)^{3/2}}{\varepsilon^{1/2}},
\end{equation}
where $\beta$ is a dimensionless constant, depending on material constants. Using the numerical values from the first-principles calculations \cite{mobility}  ($Y_x={}$1.31 eV/\AA$^2$, $Y_y={}$5.68 eV/\AA$^2$, $\varkappa_x={}$1.29 eV, $\varkappa_y={}$5.62 eV, $g_x={}$2.11 eV, $g_y={}$0.89 eV, $m_x=0.17m_e$, $m_e$ being the free electron mass) it can be estimated that $\beta\approx0.20$ \cite{app}. Introducing
\begin{equation}
\varepsilon(E)=\beta T\left[ f\left(\frac{E}{T\beta}\right)\right]^2,
\end{equation}
we arrive at the Ricatti equation for the function $f(x)$
\begin{equation}
2f'=f^2+x.
\end{equation}
At this point we note that the inequality $\beta T\ll (q'^*)^2/2m_x\varkappa^{1/2}\approx0.65T$ holds reasonably well, what justifies the above calculations.

The last equation can be turned into a linear one by the substitution $f(x)=-2\psi'(x)/\psi(x)$. The corresponding equation reads
\begin{equation}
4\psi''(x)+x\psi(x)=0,
\end{equation}
which is the Airy equation. Thus the general solution for $f(x)$ is given by
\begin{equation}
f(x)=2^{1/3}\frac{\mathop{\mathrm{Ai}}'(-x/4^{1/3})+z\mathop{\mathrm{Bi}}'(-x/4^{1/3})}{\mathop{\mathrm{Ai}}(-x/4^{1/3})+z\mathop{\mathrm{Bi}}(-x/4^{1/3})}.
\end{equation}
Here Ai and Bi are the Airy functions and $z$ is an arbitrary complex constant, which can be determined by assuring we get correct density of states for $E\rightarrow+\infty$, or in other words $\varepsilon(E)\rightarrow-E-i0$ \cite{tail}. Using the asymptotic behavior of the Airy functions it can be checked that for $z=-i$, $f^2(x)\approx-x-i/\sqrt{x}$ for large positive $x$ which yields the correct behavior for $\varepsilon$.

\begin{figure}[!tbp]
\includegraphics[width=.5\textwidth]{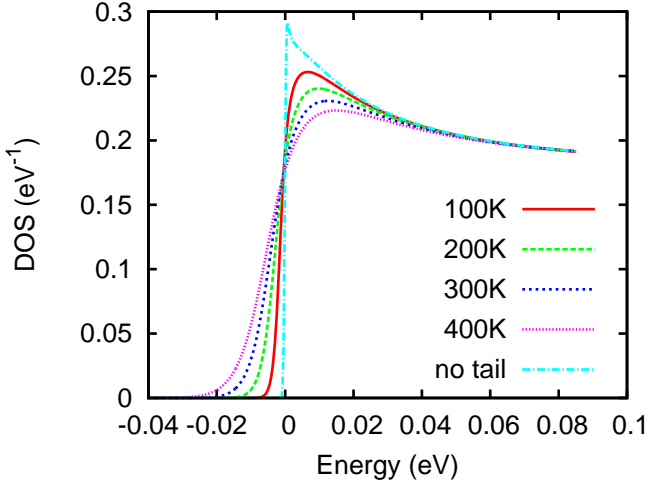}
\caption{DOS per unit cell per spin for $T=100,200,300$ and 400 K. For reference, the DOS corresponding to bare dispersion without the inclusion of the flexuron tail is shown. The Van Hove singularity in the latter manifests the aforementioned quasi-one-dimensionality of the holes in black phosphorus. It is to a large extent smeared by the flexural modes.}
\label{DOS_plot}
\end{figure}

\begin{figure}[t]
\includegraphics[width=.5\textwidth]{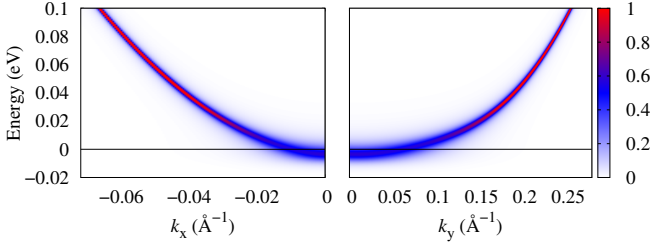}
\caption{Spectral function along $k_x$ and $k_y$ axis for $T=300$ K. Note the different scale for the two plots.}
\label{spectr}
\end{figure}

Now it is straightforward to calculate the DOS:
\begin{equation}
\rho(E)=-\frac{1}{\pi}\int\frac{d^2k}{(2\pi)^2}\Im\frac{1}{-\beta Tf^2\left(\frac{E}{\beta T}\right)-\epsilon_{\mathbf{k}}},
\end{equation}
with $\epsilon_{\mathbf{k}}$ taken from the {\it ab initio} $GW$ calculations \cite{Rudenko2015}. The result for different temperature is shown in Fig.\ref{DOS_plot}. The spectral function
\begin{equation}
A({\mathbf k},E)=-\frac{1}{\pi}\Im\frac{1}{-\beta Tf^2\left(\frac{E}{\beta T}\right)-\epsilon_{\mathbf{k}}}
\end{equation}
for room temperature is shown in Fig. \ref{spectr}.

According to the results presented on these figures the effect of flexural phonons on the density of states near the top of the valence band (the bottom of the band in hole representation) is quite noticeable and can be probed by conventional spectroscopic tools like angle-resolved photoemission or scanning probe microscopy and other tunneling methods. Quasi-one-dimensional character of the charge carrier dispersion makes hole-doped black phosphorus the best candidate to study these effects. Such experiments would be of a great interest, as an electronic probe of complicated physics of fluctuating anisotropic membranes.

The research has received funding from the European Union Horizon 2020 Programme under Grant No.~696656 Graphene Core1,
and from the Stichting voor Fundamenteel Onderzoek der Materie (FOM), which is financially supported by the Nederlandse 
Organisatie voor Wetenschappelijk Onderzoek (NWO).

\appendix
\section{Derivation of Eq.~(\ref{smallk})}
For $q'<q'^*$ we approximate a complicated pattern in ${\mathbf k'}$-integration defined by conditions ${\mathbf k'\pm q'/2}\lessgtr q'^*$ by a simpler splitting into $k'>q'^*$ and $k'<q'^*$. This is justified as the above-mentioned conditions are not strict, but rather crossovers. Here we give a detailed calculation for $k'<q'^*$ [the opposite case has been considered in the main text, Eq.~(\ref{bigk})]. 

The expression to be evaluated reads
\begin{widetext}
\begin{equation}
\label{initialintegral}
{\mathcal K}_2^{\mathrm I}(q'<q'^*)=\frac{2T^2}{(\varkappa_x\varkappa_y)^{1/4}(q'^*)^{2\eta}}\int_0^{q'^*}\frac{d^2k'}{(2\pi)^2}\frac{\left(a_x\left(k'^2_x-\frac{q'^2_x}{4}\right)+a_y\left(k'^2_y-\frac{q'^2_y}{4}\right)\right)^2}{\left|{\mathbf k'}-\frac{\mathbf q'}{2}\right|^{4-\eta}\left|{\mathbf k'}+\frac{\mathbf q'}{2}\right|^{4-\eta}}.
\end{equation}
We intruduce polar angles $\alpha$ for ${\mathbf q'}$ and $\alpha+\phi$ for ${\mathbf k'}$ as well as $x=4k'^2/q'^2$. Then the integral takes the form:
\begin{equation}
{\mathcal K}_2^{\mathrm I}(q')=\frac{T^24^{1-\eta}}{2\pi(\varkappa_x\varkappa_y)^{1/4}(q'^*)^{2\eta}q'^{2-2\eta}} \int_0^{\frac{4(q'^*)^2}{q'^2}}\!\!\!\!\!\!\!\!\frac{dx}{(x+1)^{4-\eta}}\int_0^{2\pi}\frac{d\phi}{2\pi}\frac{f(x.\alpha,\phi)}{\left(1-\frac{4x}{(1+x)^2}\cos^2{\phi}\right)^{2-\eta/2}}.
\end{equation}
Here $f(x,\alpha,\phi)$ comes from expanding the numerator in Eq.~(\ref{initialintegral}) into harmonics in $\alpha$ and reads:
\begin{multline}
f(x,\alpha,\phi)=\left[\frac14\left((x-1)^2(a_x+a_y)^2+\frac{(1+x)^2(a_x-a_y)^2}{2}\right)-\frac{x(a_x-a_y)^2}{2}\cos^2{\phi}\right]+{}\\
\left[-\frac{1}{2}(a_x^2-a_y^2)(x^2-1)+(a_x^2-a_y^2)x(x-1)\cos^2{\phi}\right]\cos{2\alpha}+{}\\
\left[\frac{(a_x-a_y)^2(1+x)^2}{8}-\frac{(a_x-a_y)^2(1+2x)x}{2}\cos^2{\phi}+(a_x-a_y)^2x^2\cos^4{\phi}\right]\cos{4\alpha}.
\end{multline}
Note that only the first term in the first brackets survives in the isotropic case. Now, using
\begin{equation}
\int_0^{2\pi}\frac{d\phi}{2\pi}\frac{1}{\left(1-\frac{4x}{(1+x)^2}\cos^2{\phi}\right)^{n-\eta/2}}={}_1F_2\left(\frac12,n-\frac{\eta}{2},1,\frac{4x}{(1+x)^2}\right)\equiv F_n,
\end{equation}
where ${}_1F_2$ is the hypergeometric function, we obtain
\begin{equation}
\int_0^{2\pi}\frac{d\phi}{2\pi}\frac{\cos^2{\phi}}{\left(1-\frac{4x}{(1+x)^2}\cos^2{\phi}\right)^{2-\eta/2}}=\frac{(1+x)^2}{4x}(F_2-F_1),
\end{equation}
and
\begin{equation}
\int_0^{2\pi}\frac{d\phi}{2\pi}\frac{\cos^4{\phi}}{\left(1-\frac{4x}{(1+x)^2}\cos^2{\phi}\right)^{2-\eta/2}}=\frac{(1+x)^4}{16x^2}(F_2-2F_1+F_0).
\end{equation}
This leads to 
\begin{multline}
{\mathcal K}_2^{\mathrm I}(q')=\frac{T^24^{1-\eta}}{2\pi(\varkappa_x\varkappa_y)^{1/4}(q'^*)^{2\eta}q'^{2-2\eta}} \int_0^{\frac{4(q'^*)^2}{q'^2}}\!\!\!\!\!\!\!\!\frac{dx}{(x+1)^{4-\eta}}\biggl\{\rule{0pt}{3.8ex}\frac{(a_x+a_y)^2}{4}(x-1)^2F_2+\frac{(a_x-a_y)^2}{8}(1+x)^2F_1+{}\\
\cos{2\alpha}\frac{(a_x^2-a_y^2)(x^2-1)}{2}\left[\frac{(x-1)}{2}F_2-\frac{(x+1)}{2}F_1\right]+\cos{4\alpha}\frac{(a_x-a_y)^2(1+x)^2}{8}\left[\frac{(x-1)^2}{2}F_2-x^2F_1+\frac{(1+x)^2}{2}F_0\right]\biggr\}
\end{multline}
Expanding the expressions in brackets for $x\rightarrow\infty$ yields $$\frac{(x-1)}{2}F_2-\frac{(x+1)}{2}F_1\sim\frac{1-\eta/2}{x}\qquad\mbox{ and }\qquad \frac{(x-1)^2}{2}F_2-x^2F_1+\frac{(1+x)^2}{2}F_0\sim\frac{-2\eta+\eta^2}{8x^2},
$$
\end{widetext}
which means that the corresponding parts of the integrand decrease at least as $x^{\eta-3}$ for large $x$, so the integrals converge well and can be extended to infinity. The rest of the integrand decreases as $x^{\eta-2}$, so the convergence for $\eta=0.85$ is quite poor. We add and subtract $\frac{(x+1)^2}{(x-1)^2}$ to $F_2$ in the first term and a unity to $F_1$ in the second term, extend the $x$-integral to infinity for well-converging parts, and keep the upper limit for the integral of $(x+1)^{\eta-2}$. Evaluating the integrals with hypergeometric functions numerically and eventually adding contribution from Eq.~(\ref{bigk}), we obtain Eq.~(\ref{smallk}). Note that the subleading, $q'$-independent term stemming from the integration over $k'<q'^*$ has the same dependence on anisotropic material constants as the contribution from $k'>q'^*$ [Eq.~(\ref{bigk})].

\section{Estimation of $\beta$}
Here we give details of the evaluation of the integral appearing in Eq.~(\ref{sigma_2}) for $\varepsilon\ll q^2/2m_x$. First we take the $\alpha$-integral, in which we keep only the leading term in $1/\varepsilon$. The denominator in the integrand has a sharp minimum at $\cos{\alpha}=0$ for small $\varepsilon$, so we can replace $\cos{2\alpha}$ by $-1$ and $\cos{4\alpha}$ by $+1$ in the numerator. Then the $\alpha$-integral is evaluated straightforwardly for both $q'>q'^*$ and $q'<q'^*$ using the identity in Eq.~(\ref{cosint}), and neglecting 1 compared to $q'^2/(2m_x\varkappa_x^{1/2}\varepsilon)$. This yields the $\varepsilon^{-1/2}$ behavior seen in Eq.~(\ref{eqsigma}). The $q'$-integrals for both regions are now trivial. To obtain $\beta$, we now have to collect all contributions and substitute $q'^*$ (the latter gives an extra $T^{-1/2}$-dependence). The results for $q'<q'^*$ are
\begin{widetext}
\begin{equation}
\beta^{3/2}\longleftarrow\left[\frac{4^{1-\eta}(2.14a_x^2+3.22a_y^2+1.00a_xa_y)}{2\eta-1}-\frac{\eta(3a_x^2+2a_xa_y+3a_y^2)}{8(1-\eta)}\right]\sqrt{\frac{16\pi}{3(Y_xY_y)^{1/2}}}\left(\frac{\varkappa_x}{\varkappa_y}\right)^{1/8}\frac{\sqrt{2m_x}}{4\pi^2}
\end{equation}
and for $q'>q'^*$:
\begin{equation}
\beta^{3/2}\longleftarrow\frac{3a_x^2-2a_xa_y+(3+8L)a_y^2}{8} \sqrt{\frac{16\pi}{3(Y_xY_y)^{1/2}}}\left(\frac{\varkappa_x}{\varkappa_y}\right)^{1/8}\frac{\sqrt{2m_x}}{4\pi^2}.
\end{equation}
\end{widetext}
Here $L$ is some numerical constant of the order of unity that comes from the logarithms in Eq.~(\ref{bigq}). If we use that expression, then $L=1+\ln{\gamma}$. Making any corrections here is meaningless, as the formal derivation of Eq.~(\ref{bigq}) implied $q'\gg q'^*$ and here we have to deal with $q'\approx q'^*$, so just ignoring the logarithm seems to be reasonable. Furthermore, $a_y$ is about 5 times smaller than $a_x$ as can be seen from the values of the material constants given in the main text. So, luckily, calculating any corrections to $L$ is irrelevant for our work and in our calculations we replaced it with 2.

Finally, substituting all material constants we obtain $\beta\approx0.20$.


\end{document}